# Laser acceleration of a thin, inflated layer of heavy material by the radiation pressure applied to a self-generated, imperfect plasma mirror.

C. STRANGIO[1] and A. CARUSO[2]
[1]Associazione EURATOM-ENEA sulla Fusione ENEA-Frascati, Italy
(strangio@frascati.enea.it)
[2]UKE (Università Kore - Enna, Italy)

**Abstract.** The production of energetic (multi-GeV) heavy ion beams by acceleration of ultra-thin foils through the application of radiation pressure to a self-generated, imperfect plasma mirror (photon absorption probability $\eta$ finite) is studied. To evaluate the foil dynamics a relativistic model was developed for a constant and relativistic invariant value of the phenomenological parameter $\eta$. The achievable efficiency of kinetic energy transfer to the matter has been evaluated as function of the parameters involved ($\eta$, the aimed average foil velocity in unit of the light speed $\beta$, etc.). The expected collimation degree for the generated ion beams, the associated energy range, the self-consistency of the model in view of the $\eta$ finite value and the survival to R-T instability were evaluated for initially thin material disks.

Energetic (tens GeV, hundreds MeV/nucleon), intense heavy ion beams are of interest for thermonuclear fusion applications [1] (e.g. fast ignition [2,3] and the injected entropy approaches [4.5]).
Energetic (100 kJ), short (ps) laser pulses can be used to create by radiation pressure the ion beams needed for the mentioned applications. According to the theory presented in this paper, to get the result the irradiation of thin solid disks with diameters of the order of hundred µm and mass of several ng is appropriate. Thus for these applications a focusing in the near field is sufficient, tight focusing mode and ultra-short pulses (e. g. fs) being unnecessary. The reference laser energy is comparable to the design value associated to the short pulse of the HiPER installation whereas the pulse duration is somewhat shorter (see [6]).

The acceleration regime described in the following was suggested by the remarkable electro-dynamic effects detected in the FIGEX experiment [7 - 10], where advanced targets, designed according to the *Controlled Amount of Mass* mode and the *Causally Isolated Target* concepts, were irradiated in the near field by high-energy, high-contrast laser pulses. In that experiment most of the ions (≈70%) were detected in the forward direction around the normal, within a cone of ≈40° FWHM. The condition of modest light absorption (absorbed fraction $\eta$=0.15) was necessary to make self-consistent all the results of the experiment since the relativistic effects on the ion flow were found modest [5,7]. Thus the ionic net forward momentum was explained as due to electro-dynamic effects (laser radiation pressure) that can overcome the thermo-kinetic ones in this regime.

The acceleration theory for a thin foil will be described in the following and systematically applied throughout the paper to the reference design of Table 1 where a heavy material solid disk featured by a high aspect ratio (*diameter/thickness*>>1) is taken as a target. In the model here considered the laser pulse impinges along the normal to the target releasing up to the time $t$ the fluence $\Gamma(t)$ (*energy/unit surface*) at a power density $\Phi_{laser}(t)=d\Gamma/dt$. The layer, in its own reference system, is assumed in a quasi-neutral quasi-equilibrium under the actions of the internal pressure and of the pseudo-gravity field due to the acceleration. This implies that $\lambda_D<<\Delta$, where $\Delta$ and $\lambda_D$ are the target thickness and the Debye length at the generic time $t$. The thin disk is assimilated to a 1-D system accelerated as a whole by the laser radiation pressure applied to the irradiated side. The adopted motion description includes the relativistic corrections to extend the model to heavy ion energies of interest for thermonuclear applications ($0.2<\beta<0.5$). The photons absorbed by the target (probability $\eta$) transfer their energy ($h\nu$) and momentum ($h\nu/c$) to the electrons while the reflected fraction (1-$\eta$) transfers to the electrons a momentum depending on the velocity of the irradiated surface due to the frequency shift after the reflection. The thermal effects on the hydrodynamic flow induced by the fast electron heating are introduced, as a correction, through the mentioned *phenomenological* constant factor $\eta$ (with values eventually deduced from the experiments). For the electronic and ionic particle densities ($n_e$ and $n_i$, in the moving layer system), the quasi-neutrality condition implies $Z_i n_i \approx n_e$ ($Z_i$ is the ionic charge) *in the bulk of the target*. The electrons are heated to a temperature $T_h$ (of the order of several ten MeV) [11] by the absorbed fraction of the laser energy and move in a collisionless regime at velocities of the order of the light speed $c$. Since $\lambda_D<<\Delta$ the fraction of escaping electrons is small. The electronic round-trip through the target has a value of $\approx 2\Delta/c$, much smaller than the laser durations ($\tau_{laser}$) here considered. The light skin depth penetration $\delta$ into the matter is $\approx \lambda_D$, as deduced by $\delta \approx \lambda_D c/V_{th}$, where $V_{th}$ is the electronic thermal velocity.

| *Solid disk initial parameters* | |
|---|---|
| Initial geometry | Diameter/Thickness >>1 |
| Radius $R_o$ | 50 (μm) |
| Thickness $\Delta_o$ | 57 (nm) |
| Material | Bi |
| Density | 9.8 (g/cm$^3$) |
| *Laser beam and focusing* | |
| Energy | 100 (kJ) |
| Wavelength | 0.527 (μm) |
| Duration | 0.5 (ps) |
| Spot radius | 50 (μm) |
| Power density | 2.5×10$^{21}$ W/cm$^2$ |
| *Laser generated ion beam* | |
| Aimed MeV/nucleon | 72 ($\beta$=0.372) |
| $\eta$ | 0.15 |

Table 1 – The reference laser energy is comparable to the design value associated to the short pulse of the HiPER installation whereas the pulse duration is somewhat shorter (see RAL-TR-2007-2008).

On the irradiated side of the target the electrons are pushed forward by the electro-magnetic field into the ionic gas up to a distance of the order of $\lambda_D \approx \delta$ (positive sheet) while, at the target rear side, a negative sheet (thickness about $\lambda_D$) arises due the hot electronic gas restrained by the a self-consistent ion electrostatic field [8]. In the bulk of the material the ions are acted by the electric field $E$ due to the quasi-neutral coupling (Fig. 1). This field can be estimated through the electron gas equilibrium equation, written in the system co-moving with the accelerated layer, as $E \times \Delta \approx p_{eb}/(eZ_i n_i) \approx T_h/e$, where $p_{eb}$ is the electronic pressure at the boundary illuminated by the laser light.

In the laboratory reference frame the equation of the motion for the layer takes into account the effects of the radiation pressure $p_{tot} = p_{abs} + p_{refl}$ where the two contributions, $p_{abs}$ and $p_{refl}$, respectively correspond to the absorbed and reflected fraction of photons.

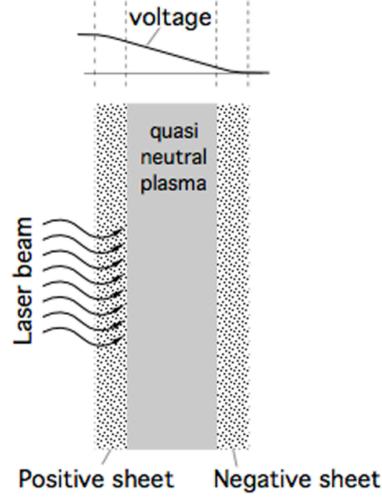

Fig.1- Plasma layer accelerated by radiation pressure: coupling voltage distribution.

The equation for a layer of mass $m$ per unit surface pushed by the radiation pressure $p_{tot}$ can be integrated analytically [10] to yield the equation $\Gamma_{laser}(t) = \Psi(\beta,\eta) \times mc^2$, where

$$\Psi(\beta,\eta) = \frac{1}{2}(\frac{\sqrt{1-\beta^2}}{1-\beta}-1) + \frac{\eta}{4\sqrt{1-\eta}}\arctan\{\frac{2\sqrt{1-\eta}\,[\beta(2-\eta)+\eta(1-\sqrt{1-\beta^2})]}{4\sqrt{1-\beta^2}(1-\eta)+\beta(2-\eta)\eta+\eta^2}\}$$

and the hydrodynamic efficiency can be defined as the ratio of the kinetic energy $K = mc^2[(1-\beta^2)^{-1/2}-1]$ to $\Gamma_{laser}$ namely $\eta_{hydr} = [(1-\beta^2)^{-1/2}-1]/\Psi(\beta,\eta)$.

For the reference quantities of Table 1, it results $\Psi(0.372,0.15)=0.254$ and $\Gamma_{laser}=1.27$ GJ/cm$^2$.

Another interesting parameter is the ratio $\eta_{hydr}/\eta$ giving an estimate of the relative importance of the hydrodynamic forward directed energy flow to the thermal one that in the layer reference system drives an intrinsically symmetric thermal expansion when the radiation pressure ends (at the time $t=\tau_{laser}$). The efficiency $\eta_{hydr}$ and the ratio $\eta_{hydr}/\eta$ are represented in Fig. 2, a) and b)

Values of these quantities for the quoted reference case are $\eta_{hydr}=0.304$, $\eta_{hydr}+\eta=0.454$ and $\eta_{hydr}/\eta=2.03$. These results show that the effect due to the forward motion induced by the radiation pressure is the most important process. It has to be noted that $\beta=0.372$ and $\eta=0.15$ correspond to a 15 GeV Bi ion beam (72 MeV/nucleon) and to an absorption coefficient consistent with the FIGEX experimental results (see Table 1).

It is important now to check the self-consistency of the previous model where the layer is considered, in its own reference system, in equilibrium under the effects of the internal pressure and the pseudo-gravitational field created by the acceleration. At the interface material-radiation the electronic pressure $p_{eb} = n_e T_h = n_i Z_i T_h$ must be balanced by the radiation pressure evaluated in the co-moving system by $p_{eb} = (2-\eta)\phi_{rest}/c$, where $\phi_{rest}$ is laser power density in the comoving system [10]. The accelerated layer thickness $\Delta$ can be estimated by considering a quasi-neutral, isothermal equilibrium in this system. This quantity depends on the isothermal sound velocity $c_s = \sqrt{\eta \Gamma_{rest}/m}$ ($\Gamma_{rest}$ is the fluence in the comoving system), on the applied radiation pressure and on the initial areal mass density $m$ (a conserved quantity). The initial solid matter density $\rho_o$ and the layer thickness $\Delta_o$ are expected to enter in the combination $m=\rho_o\Delta_o$, when the conditions $\Delta >> \Delta_o$ and $\rho << \rho_o$ hold for the heated layer ($\rho \Delta = \rho_o \Delta_o = m$). The only combination having the dimension of a length is $\Delta = m c_s^2 / p_{eb}$. The quantity $\Delta$ represents exactly the height scale of an isothermal atmosphere under the pseudo-gravity $g = p_{eb}/m$. At the end of the acceleration, for the reference situation of Table 1 results $\Delta$=14 µm. This model holds until the layer remains opaque to the laser radiation due to the absorption and reflection processes. A sufficient condition can be $\rho_c < \rho \approx m/\Delta$, where the critical density $\rho_c$ is calculated for the red-shifted laser radiation in the layer reference frame.

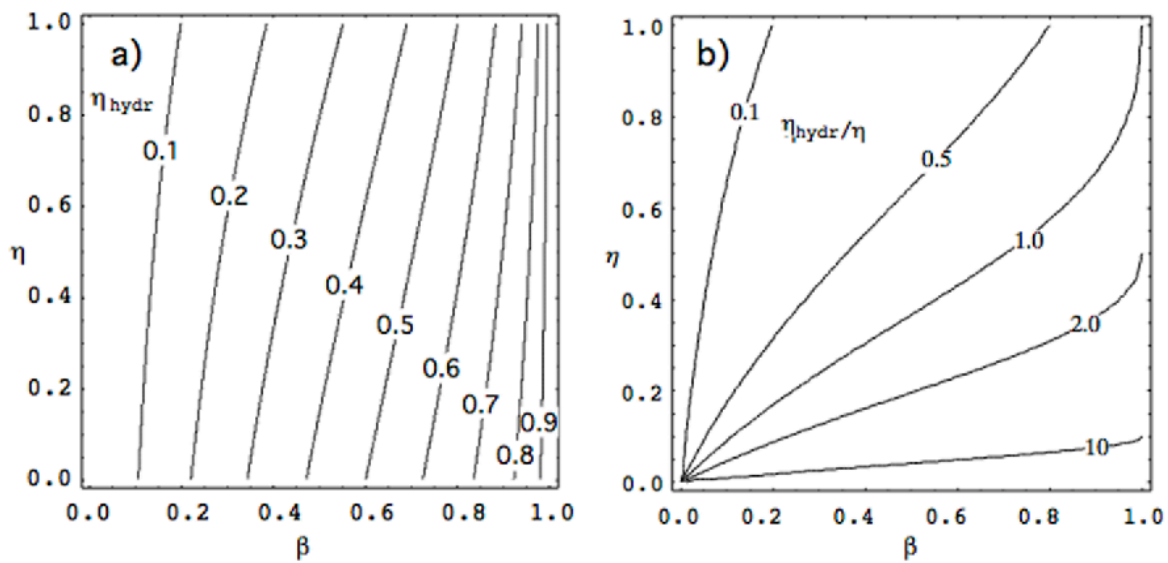

Fig. 2 – The Hydrodynamic efficiency $\eta_{hydr}$ and the efficiencies ratio $\eta_{hydr}/\eta$ both as a function of the aimed $\beta$, and $\eta$.

Another relevant issue is that of the Rayleigh-Taylor instability effects. In this case bubble propagation into the layer can make laser radiation filaments to propagate inside the matter producing a sort of R-T driven ablation. An estimate of the relevance of the R-T process can be obtained according to the model described by Youngs [12, 13] where the effect of long-wavelength initial perturbations are assumed negligible and the thickness of the bubble layer, $h(t)$, are evaluated by dimensional methods with a normalization arising from experiments and numerical simulations.

For the layer motion here considered the thickness $h(t)$ was assumed to satisfy the equation $d^2h/dt^2 = 2\alpha g$ where $g = p_{eb}/m$ and $\alpha = 0.06 \div 0.07$. This equation was integrated and the ratio $h/\Delta$ evaluated.

The consistency analysis shows, for the reference case of Table 1, that transparency does not occur up to the end of the laser pulse (at the time $\tau_{laser}$), whereas about 80% of the layer remains free of the R-T bubbles (estimate performed for $\Phi_{laser} \approx$ const during $\tau_{laser}$).

From the previous model it is possible to infer, at a semi quantitative level, the degree of collimation of the generated ion beam. This can be done by noting that when the laser pulse ends at $\tau_{laser}$, also the radiation pressure ends and in the co-moving system a roughly symmetric expansion starts, forward and backward. The gas expansion results somewhat collimated around the direction perpendicular to the front and back cylinder surfaces when, at the time $t = \tau_{laser}$, the layer thickness $\Delta$ results smaller than the target radius $R$. Under this conditions longitudinal velocities of the order of $v_{\parallel} \approx c_s$ are achieved, larger than the transverse ones $v_{\perp}$ according to the ratio $v_{\parallel}/v_{\perp} \approx 2R/\Delta$ [5]. The corresponding half-aperture is $\theta \approx \arctan(\Delta/2R)$ both in the forward and in the backward beams.

It has to be noted that, for $\beta_s = c_s/c < \beta$, in the laboratory reference frame after the expansion most of the ions remain projected in the forward direction producing an ion-beam with efficiency of the order of $\eta_{tot} \approx \eta_{hydr} + \eta$. This situation is verified for the reference case for which $\eta_{tot} \approx 0.45$.

The forward and the backward moving ions are emitted [14] within the half-angles $\theta_{forw} = \arctan[\beta_s \sqrt{1-\beta^2} \sin\theta /(\beta_s \cos\theta + \beta)]$ and $\theta_{backw} = \arctan[\beta_s \sqrt{1-\beta^2} \sin\theta /(\beta - \beta_s \cos\theta)]$. For the reference case ($R \approx 50\,\mu m$, $\Delta \approx 14\,\mu m$) it is found $\theta_{forw} = 2.3°$ (0.04 rad), $\theta_{backw} = 6.4°$ (0.11 rad) and $\theta_{backw}/\theta_{forw} \approx 3$. The ion bulk has a speed in the interval $0.26 < \beta < 0.51$, corresponding to energies per nucleon $E_n$ (in MeV) ranging in $21 < E_n < 155$. In practice it is to be expected a forward moving, quite collimated beam of high-energy particles surrounded by a halo filled by particles of somewhat lower energy.

For a target shaped as a spherical-cup layer of radius $f$ it is possible to expect at the center of curvature a spot of radius $f \times \theta_{forw}$ for the high-energy group. This spot will be surrounded by a halo of radius $f \times \theta_{backw}$ filled by the lower energy ions. As an example, for $f = 500\,\mu m$, using the value of $\theta_{forw} = 0.04$ rad ($\approx 2.3°$) previously obtained, it is found $f \times \theta_{forw} \approx 20\,\mu m$, a value of interest for fast ignition applications. Values of interest for the injected entropy approach to IFE would be already obtained for $f = 0.5$ cm.

In conclusion, the generation of fast heavy ions (e.g. Bi, U ions, tens GeV) by the irradiation of ultra-thin targets (large diameter to thickness ratio) with short laser pulses (e.g. 0.5 ps) at power densities of some $10^{21}$ W/cm$^2$ was studied having in mind applications like those to the inertial confinement.

The regimes under consideration can be generated by laser radiation pressure applied to a self-generated plasma mirror in regimes of low power absorption (e.g. $10 \div 15$ %, imperfect plasma mirror). The radiation pressure induces the most important hydrodynamic effects and makes possible the efficient generation ($30 \div 45\%$) of all-forward ejected high-energy ions ($20 \div 150$ MeV/nucleon) with the highest energy part of the spectrum well collimated (within a half angle as small as $\theta_{forw} \approx 2°$ for the reference situation of Table 1). The process can survive to the expansion process induced by the finite light absorption and to bubble penetration due to Rayleigh-Taylor instability for a time adequate to get the sought energies. Moreover it has been shown that using as ion source a spherical cup with radius of curvature $f$, a spot radius of about $f \times \theta_{forw}$ can be obtained on a target set at the distance $f$ from the source. This size is compatible with the applications to IFE for reasonable values of $f$. Also the spreads in velocity are amply compatible with the on-target pulse durations required for both the IFE applications, namely fast ignition by heavy-ions and injected entropy.

## References

[1] CARUSO A, STRANGIO C., Studies on non conventional high-gain target design for ICF. *Laser and Particle Beams*, **19**, 295–308 (2001).

[2] CARUSO A. AND PAIS, V.A., The ignition of dense DT fuel by injected triggers. *Nucl. Fusion* **36** 745-57 (1996).

[3] CARUSO A. , PAIS V. A., Effects of the injected trigger pulse focusing and timing on the ignition and gain of dense static, or imploding DT fuel. *Physics Letters A* 243 319-324 (1998).

[4] CARUSO A., STRANGIO C., The injected entropy approach for the ignition of high gain targets by heavy ions beams or incoherent x-ray pulses *IFSA99 Proceedings*, 88-93 (1999).

[5] CARUSO A., STRANGIO C., Ignition of high gain targets by entropy injection. *Laser and Particle Beams,* **18**, 35–47 (2000).

[6] RAL-TR-2007-2008.

[7] STRANGIO C., CARUSO A., NEELY D., ANDREOLI, P.L., ANZALONE R., CLARKE R., CRISTOFARI G., DEL PRETE E., DI GIORGIO G., MURPHY C., RICCI C., STEVENS R., TOLLEY M., Production of multi-MeV per nucleon ions in the controlled amount of matter mode (CAM) by using causally isolated targets. *Laser and Particle Beams,* **25**, 1–7 (2007).

[8] STRANGIO C., CARUSO A. Comparison of fast ions production modes by short laser pulses. *Laser and Particle Beams*, **23**, 33–41 (2005).

[9] STRANGIO C., CARUSO A., ICF applications of fast ions generated by focusing short laser pulses on ultra-thin causally isolated targets. *Proceeding of IFSA 2007, J. Phys.: Conf. Ser.* **112** (2008) 042043 (4pp).

[10] STRANGIO C., CARUSO A., AGLIONE M. Studies on possible alternative schemes based on two-laser driver for Inertial Fusion Energy applications *Laser and Particle Beams*, **27** 303-309, (2009).

[11] HATCHETT, S.P., BROWN, C.G., COWAN, T.E., HENRY, E.A.,JOHNSON, J.S., KEY, M.H., KOCH, J.A., LANGDON, A.B.,LASINSKI, B.F., LEE, R.W., MACKINNON, A.J., PENNINGTON,D.M., PERRY, M.D., PHILLIPS, T.W., ROTH, M., SANGSTER,T.C., SINGH M.S., SNAVELY, R.A., STOYER, M.A., WILKS,S.C., YASUIKE, K., Electron, photon, and ion beams from the relativistic interaction of Petawatt laser pulses with solid targets. *Phys. Plasmas* **7**, 2076–2081, (2000).

[12] YOUNGS D. L., Numerical simulation of turbulent mixing by Rayleigh-Taylor instability *Physica* **12D**, 32-44 (1984).

[13] YOUNGS D. L., Modeling turbulent mixing by Rayleigh-Taylor instability *Physica* **D 37**, 270-287 (1989).

[14] LANDAU L.D., LIFSCHITZ E. M. The classical theory of fields *Course of Theoretical Physics*, Vol. II, Pergamon
  Press (Oxford-London-Paris-Frankfurt, 1962).